\def\Herb{$\rm ZnCu_3(OH)_6Cl_2$}

\def\cm{cm$^{-1}$}
\documentclass[aps,prb,twocolumn,showpacs,superscriptaddress,am,floatfix]{revtex4-1}
\usepackage{graphicx}
\usepackage{amsmath}
\usepackage{amssymb}
\usepackage{color}
\begin{document} 
\title{Lattice dynamics  in the spin-1/2 frustrated kagome compound
herbertsmithite}
\author{Ying Li*}  
\affiliation{Department of Applied Physics and MOE Key Laboratory for Nonequilibrium Synthesis and Modulation of Condensed Matter, School of Science, Xi'an Jiaotong University, Xi'an 710049, China}
\affiliation{Institut f\"ur Theoretische Physik, Goethe-Universit\"at Frankfurt,
Max-von-Laue-Strasse 1, 60438 Frankfurt am Main, Germany}
\author{A. Pustogow*} 
\affiliation{1.~Physikalisches Institut, Universit\"{a}t
Stuttgart, Pfaffenwaldring 57, D-70569 Stuttgart Germany}
\author{M. Bories}
\affiliation{1.~Physikalisches Institut, Universit\"{a}t
Stuttgart, Pfaffenwaldring 57, D-70569 Stuttgart Germany}
\author{P. Puphal}
\affiliation{Physikalisches Institut, Goethe-Universität Frankfurt, 60438 Frankfurt am Main, Germany}
\author{C. Krellner}
\affiliation{Physikalisches Institut, Goethe-Universität Frankfurt, 60438 Frankfurt am Main, Germany}
\author{M. Dressel}
\affiliation{1.~Physikalisches Institut, Universit\"{a}t
Stuttgart, Pfaffenwaldring 57, D-70569 Stuttgart Germany}
\author{Roser Valent\'i}
\affiliation{Institut f\"ur Theoretische Physik, Goethe-Universit\"at Frankfurt,
Max-von-Laue-Strasse 1, 60438 Frankfurt am Main, Germany}
%

\date{\today}
\begin{abstract} {We investigate the lattice dynamics in the
	spin-1/2 frustrated kagome compound
        herbertsmithite ZnCu$_3$(OH)$_6$Cl$_2$ by a combination of
	infrared spectroscopy measurements and
	{\it ab initio} density functional theory  
	calculations, and provide an unambiguous assignment
	of infrared-active lattice vibrations involving in-plane and out-of-plane
	atom displacements in the kagome layers.
	Upon cooling, non-thermal red-shifts and broadening appear specifically for modes
	that deform the kagome layer or affect the Cu-O-Cu bond angles, thus
	creating pronounced modifications of the antiferromagnetic
	exchange coupling.  Our results indicate 
	the presence of a strong magnetoelastic coupling to the 
	spin system.
	We discuss the effects of this
	coupling and its relation
	to recent experiments reporting a global symmetry reduction
	of the kagome lattice symmetry.}
\end{abstract}
%


\maketitle
%
%

The nature of the ground state in the frustrated spin-1/2
 kagome compound ZnCu$_3$(OH)$_6$Cl$_2$ has been
	subject of intense discussion for many years \cite{Norman2016,Shaginyan2019}. Being
	a perfect realization of a kagome lattice of spin-1/2
	Cu atoms with dominant nearest-neighbor
	Heisenberg antiferromagnetic interactions, $J$ $\approx$ 180 K,\cite{Jeschke2013}
	this material was suggested to be a canonical candidate for
	bearing a quantum-spin-liquid (QSL) state. Indeed, the Cu$^{2+}$ magnetic moments
	do not  order \cite{Mendels2007}
	down to the lowest measured temperatures and excitations are
	dominated by a rather unconventional
	broad continuum \cite{Han2012}. While these features
	strongly indicate a QSL ground state, there is presently
	a lot of debate, both from the experimental and theoretical side,
	concerning its nature \cite{Shores2005, Helton2007, Mendels2007, DeVries2009, Imai2008, Olariu2008, Singh2007, Ran2007, Depenbrock2012, Iqbal2013, Fu2015, He2017, Liao2017}. Interestingly,
	recent studies on herbertsmithite
	have invoked different aspects of this compound, such as
	interlayer Cu/Zn antisite disorder \cite{Khuntia2019}
	and lattice effects \cite{Laurita2019} to investigate
	this question. Actually, such effects uncover a rich behavior
	in these systems. For instance the authors of Ref. \onlinecite{Laurita2019}
	identify via non-linear optical
        response experiments a subtle high-temperature
        monoclinic distortion in herbertsmithite that is suggested 
	to influence the nature of the QSL ground state. Such symmetry reduction was
	also reported in previous torque and electron-spin-resonance (ESR) studies \cite{Zorko2017}.
	Its origin and relation to the underlying spin system, however, remain unsettled.
	The knowledge on the nature
	of the vibrational modes in this system and the magnetoelastic
	coupling  would be of major help to resolve these questions.
        In fact, there has already been 
	considerable spectroscopic work on herbertsmithite 
	investigating phononic, 
	electronic and magnetic
	excitations~\cite{Sushkov2017,Pustogow2017Herbertsmithite,Wulferding2010}. 
        Specifically, recent infrared spectroscopy studies on the title compound and other frustrated systems revealed
        anomalous broadening and red shifts of vibrational modes upon
        cooling,~\cite{Sushkov2005,Sandilands2015,Sushkov2017} as opposed to
        the narrowing and hardening arising from the usual thermal effects. It was suggested that these anomalies are related to the coupling to the
	spin system.
 
	In view of these findings, we performed
 an extensive theoretical and spectroscopic study
	of the vibrational features in $\rm
	ZnCu_3(OH)_6Cl_2$. Via $ab~initio$ phonon calculations we provide a one-to-one
	assignment of phonon modes and the involved lattice
	sites. We observe non-thermal broadening and red-shift for
	low-frequency vibrations in the range of expected strong spin fluctuations.
	Notably, we identify these low-frequency vibrations to be
	related to atomic motions
        that deform the kagome layers and modify the Cu-O-Cu bond angles directly affecting
	the antiferromagnetic Cu-Cu exchange interactions and therefore influencing the
	underlying magnetism.


Single crystals of \Herb\ were prepared by hydrothermal
	synthesis~\cite{Han2011}. The optical reflectivity was measured on
	as-grown surfaces of mm-sized samples covering frequency and
	temperature ranges of 40--20\,000~\cm\ and 10--295~K, respectively, and
	combined with the VIS-UV data reported
	elsewhere~\cite{Pustogow2017Herbertsmithite}. Standard extra\-polations
	towards low and high frequencies were applied in order to determine the
	optical conductivity using the Kramers-Kronig relations. 
	
	$Ab~initio$
vibrational studies were performed using the
	crystal structure of
	\Herb\ given in the space group $R\bar{3}m$~\cite{Braithwaite2004}.
The phonon frequencies and eigenvectors were calculated by diagonalizing the
dynamical matrices using the phonopy package~\cite{Togo2008,Togo2015}. The
dynamical matrices were constructed from the force constants determined from
the finite displacements in  $2 \times 2 \times 1$
supercells~\cite{Parlinski1997} performing density functional calculations within
the Perdew-Burke-Ernzerhof parametrization of the generalized gradient
approximation~\cite{Perdew1996} and using the projector augmented wave
approximation~\cite{Bloechl1994} implemented in the Vienna package
(VASP)~\cite{Kresse1993, Kresse1996a, Kresse1996b}. The Brillouin zone for the
supercell was sampled with a $4 \times 4 \times 4~\mathbf{k}$ point mesh, and
the plane-wave cut-off was set at  520 eV. 


\begin{figure}
 \centering
 \includegraphics[width=1\columnwidth]{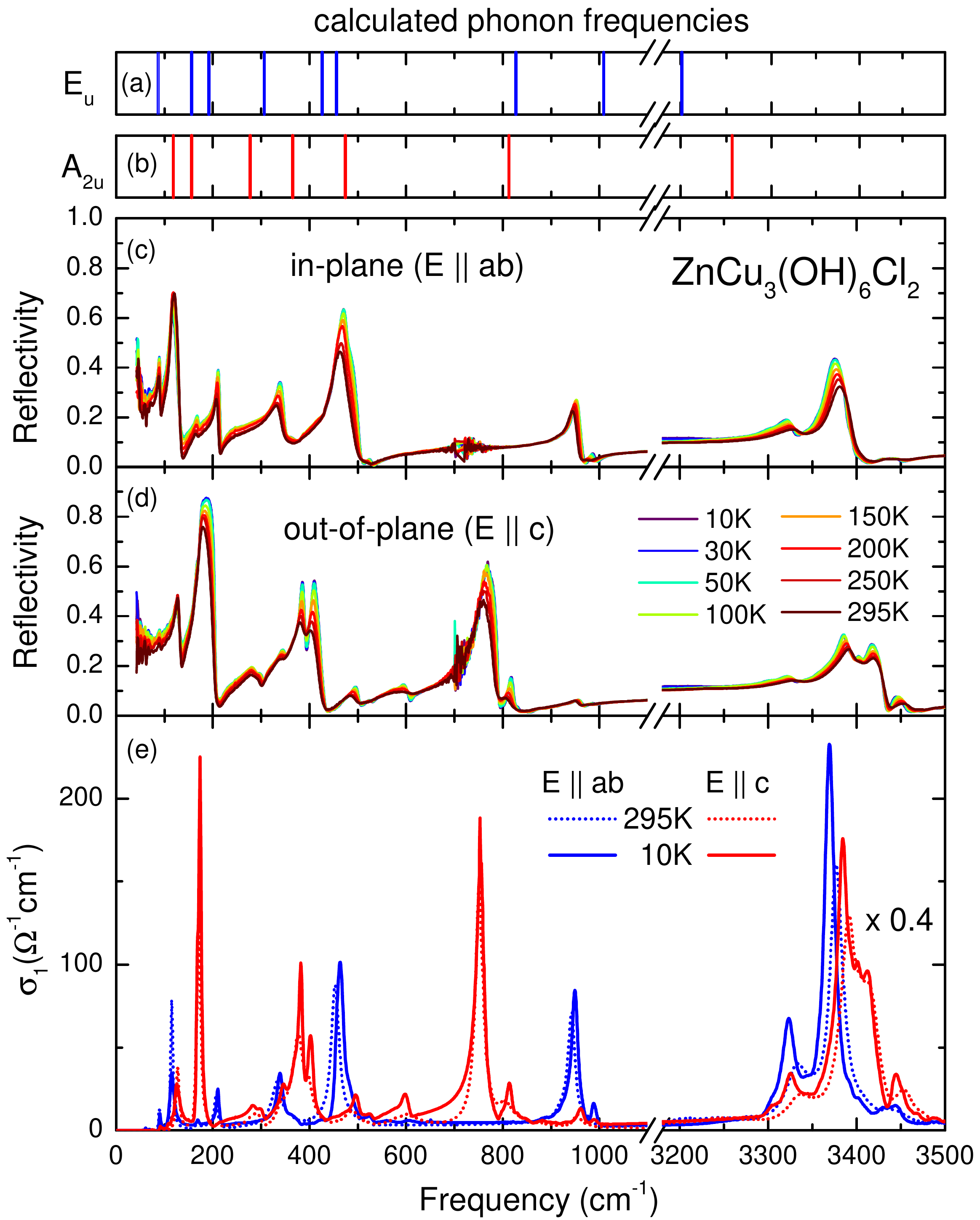}
 \caption{(a,b) Phonon calculations yield good agreement with polarized infrared spectra of \Herb. Note the $\rm E_u$ doublet at 84.8 and 88.4~\cm; all E$_{u}$ and A$_{2u}$ mode frequencies are listed in Table~\ref{tab:freq}. (c,d) The temperature-dependent in- and out-of-plane reflectivity is dominated by vibrational features in the entire infrared range. (e) The corresponding optical conductivity was calculated using the Kramers-Kronig relations. For convenience, only 295~K (dotted) and 10~K (solid lines) spectra are shown for both orientations. 
}
\label{comparison}
\end{figure}

\begin{table}[b]
\caption {The infrared-active doubly-degenerate E$_u$ modes and infrared-active out-of-plane A$_{2u}$ modes in cm$^{-1}$ for \Herb\  obtained by $ab~initio$ calculations.}
\centering
\def\arraystretch{1.1}
\label{tab:freq}
\begin{ruledtabular}
\begin{tabular}{lllllllll}
E$^1_{u}$ & E$^2_{u}$ & E$^3_{u}$ & E$^4_{u}$ & E$^5_{u}$ & E$^6_{u}$ & E$^7_{u}$ & E$^8_{u}$ & E$^9_{u}$\\
\hline
84.8  &88.4 &155.8 &191.8 &306.5 &426.1 &455.8 &826.5 &1008.7 \\
\hline
A$^1_{2u}$ & A$^2_{2u}$ & A$^3_{2u}$ & A$^4_{2u}$ & A$^5_{2u}$ & A$^6_{2u}$ & A$^7_{2u}$ &  &  E$^{10}_{u}$ \\
\hline
117.9 &155.9 &277.2 &365.1 & 473.8 &812.4 &3256.3 &  &  3201.5 \\
\end{tabular}
\end{ruledtabular}
\end{table}

\begin{figure}
 \centering
 \includegraphics[width=1\columnwidth]{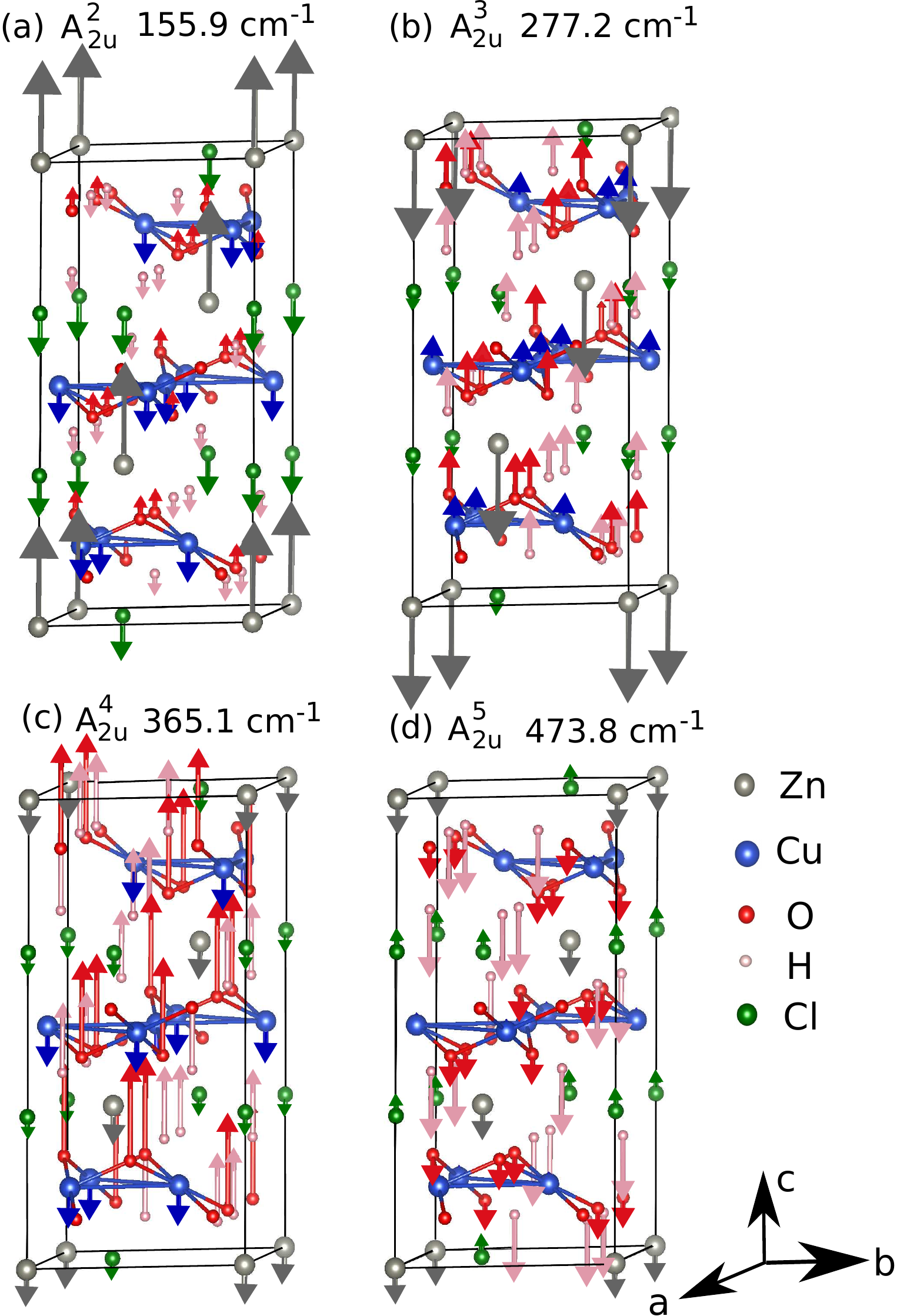}
	\caption{The $A_{2u}$ modes at (a) 155.9 cm$^{-1}$ (b) 277.2 cm$^{-1}$ (c) 365.1 cm$^{-1}$ and (d) 473.8 cm$^{-1}$ according to our {\it ab initio} calculations. The atomic motions of these infrared vibrations were projected along the crystallographic $c$-axis.}
\label{fig:a2uother}
\end{figure}

\begin{figure}
 \centering
 \includegraphics[width=1\columnwidth]{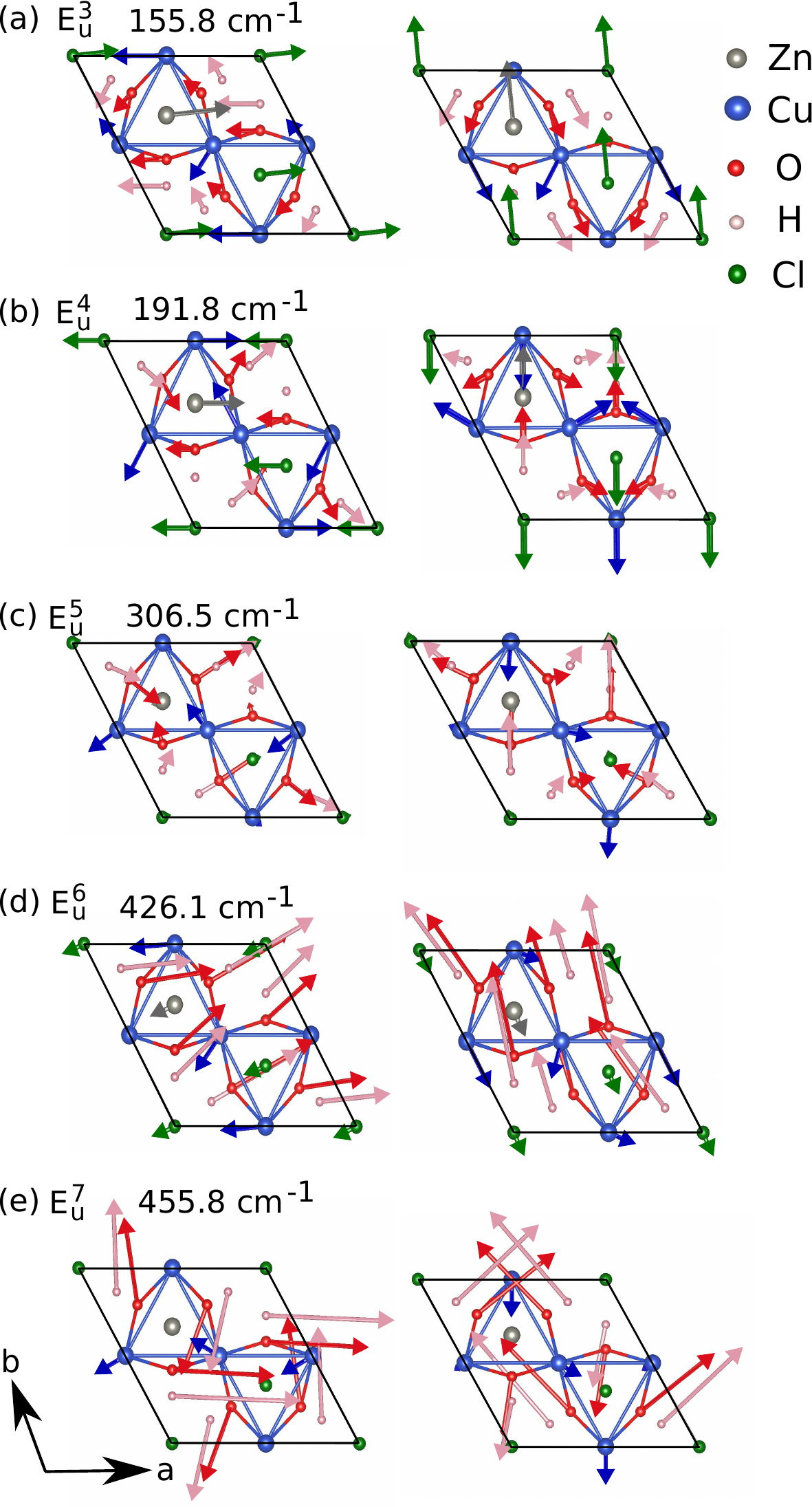}
	\caption{The doubly-degenerate $E_{u}$ infrared vibration modes at (a) 155.8, (b) 191.8, (c) 306.5, (d) 426.1, and (e) 455.8 cm$^{-1}$ according to our {\it ab initio} calculations. The arrow sizes and lengths correspond to the relative amplitudes of the modes, which are projected on the $ab$-plane.
}
\label{fig:e2other}
\end{figure}
In Fig.~\ref{comparison}(c,d) we show the polarized infrared reflectivity spectra of \Herb\ 
and
compare them to the calculated phonon mode frequencies
shown in Fig.~\ref{comparison}(a-b). The optical absorption
in herbertsmithite arises mainly from phonons since the strong Coulomb
repulsion  in this system yields a charge gap of more than
3~eV~\cite{Pustogow2017Herbertsmithite}. This is in contrast to the case
of triangular-lattice organic QSL candidates, where electronic
transitions between the Hubbard bands are the dominant contribution in the far-
and mid-infrared range~\cite{Pustogow2018,Dressel2018,Ferber2014}. While vibrational calculations have been carried out and discussed also on the molecular solids~\cite{Dressel2016,Matsuura2019}, the absence of an electronic background in \Herb\ allows for an undisturbed investigation of the coupling between magnetic excitations and the lattice.
In Fig.~\ref{comparison}(c,d)
the anisotropy and temperature 
dependence of the in-plane (electric field $E\parallel ab$) 
and out-of-plane reflectivity ($E\parallel c$)
agree well with previous reports~\cite{Sushkov2017,Pilon2013},
confirming our excellent sample quality. We plot the corresponding optical
conductivity at room temperature and at 10~K in Fig.~\ref{comparison}(e).

Herbertsmithite in the $R\bar{3}m$ structure has 18 atoms in the primitive
unit cell and $18\times 3  = 54$  phonon modes are expected.
Our calculations display all 54 phonon modes as previously classified~\cite{Sushkov2017} including 10
infrared-active doubly-degenerate $\rm E_u$ in-plane modes  	 
($E\parallel ab$, Fig.~\ref{comparison}(a)) and 7 infrared-active 
out-of-plane modes $\rm
A_{2u}$  ($E\parallel c$ Fig.~\ref{comparison}(b)). The frequencies are shown in Table \ref{tab:freq}. Full phonon modes can be found in the Supplemental Material~\footnote{See Supplemental Material which contains all 54 phonon modes for Herbertsmithite}. Our calculations agree well with our measured optical spectra (Fig.~\ref{comparison}(c-e)). Especially at low frequencies we can unambiguously
assign the experimentally observed peaks to the computational results. The atomic motions of the infrared vibrations between 150 and 500 cm$^{-1}$ are displayed in Fig. \ref{fig:a2uother} for the $A_{2u}$ modes (projected along $c$ direction) and Fig. \ref{fig:e2other} for the $E_{u}$ modes (projected onto the $ab$ plane). The total dipole moment of the $A_{2u}$ modes parallel to the $ab$ plane and for the $E_{u}$ modes along $c$ direction is zero, consistent with the polarized infrared spectra. While the lattice vibrations are dominated by Zn, Cl and Cu motions in the region 150--300 cm$^{-1}$, O and H atoms (and less pronounced also Cu) exhibit the largest amplitudes between 300 and 500 cm$^{-1}$.
In the
frequency range 350--500~\cm\, Cu-O scissoring and stretching are the dominant
motions, with oxygen atoms experiencing the largest amplitude.
For this reason, these and higher lying vibrations are most sensitive to
Cu/Zn antisite disorder which is likely the origin of the splittings (400 and
700-820~\cm\ for $E\parallel c$; 900--1000~\cm\ for $E\parallel ab$) and the additional
features with small intensity (600~\cm\ for $E\parallel c$). These
splittings may also result from the lower monoclinic symmetry as suggested in
Ref.~\onlinecite{Laurita2019}.
Above 3000~\cm\  the relative position of the out-of-plane hydrogen-related modes
compared to those measured in-plane are in line with our calculations.
We suggest that these O-H stretching
vibrations around 3300--3500~\cm\  are also affected by antisite
disorder, which explains their complex splitting: each proton is attached to an oxygen atom and feels the modified
crystallographic surroundings in a similar way.

\begin{figure}
 \centering
 \includegraphics[width=1\columnwidth]{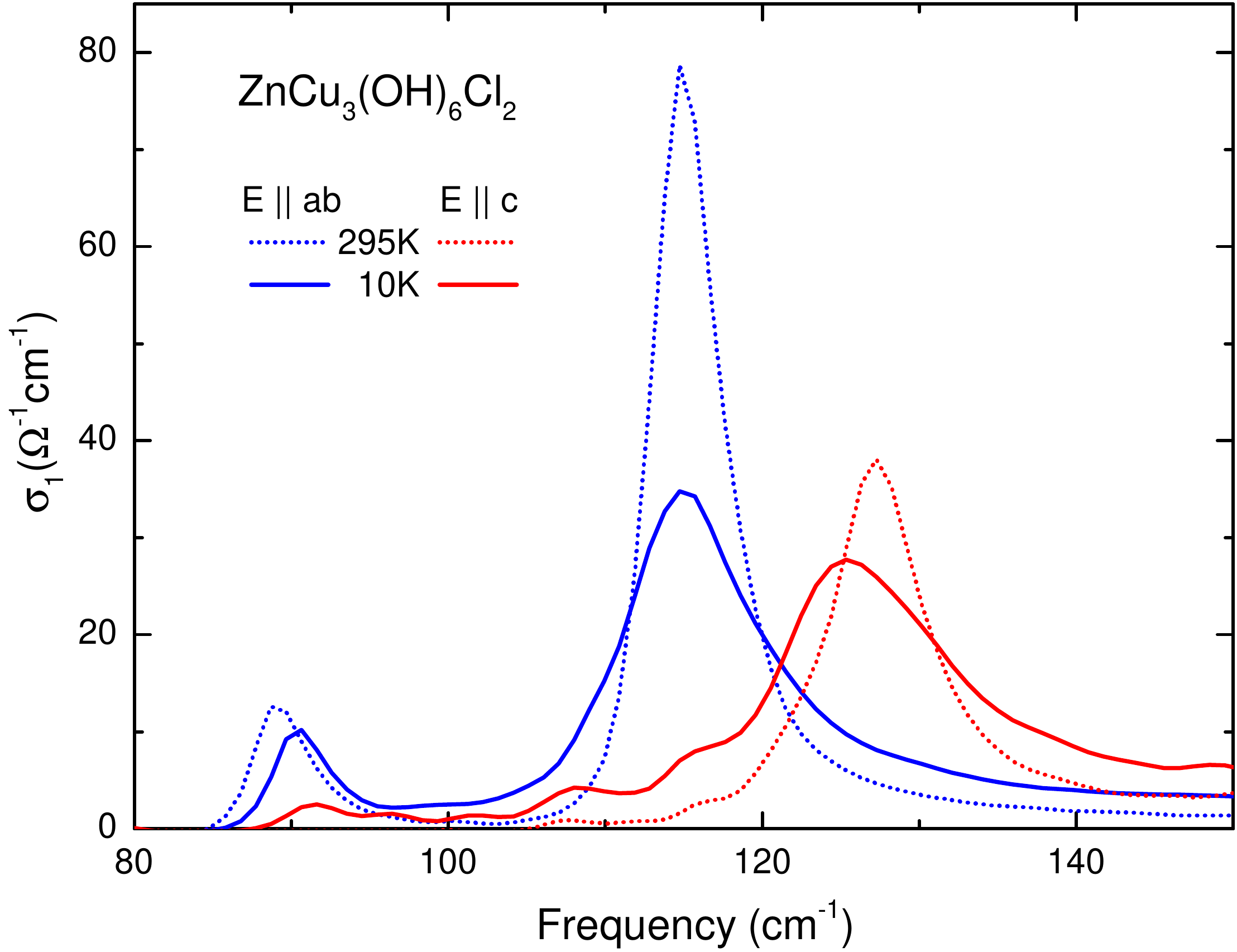}
 \caption{Detailed view on the measured 
	lowest-energy infrared vibrations of \Herb. The
	magneto-elastic coupling is expected to be
	strongest for the modes at 115~\cm\ ($E\parallel ab$) and 125~\cm\ ($E\parallel c$),
	as expressed by the pronounced broadening and red-shift upon cooling.}
\label{Zn_low-energy-sketch}
\end{figure}

\begin{figure*}
\includegraphics[width=\textwidth]{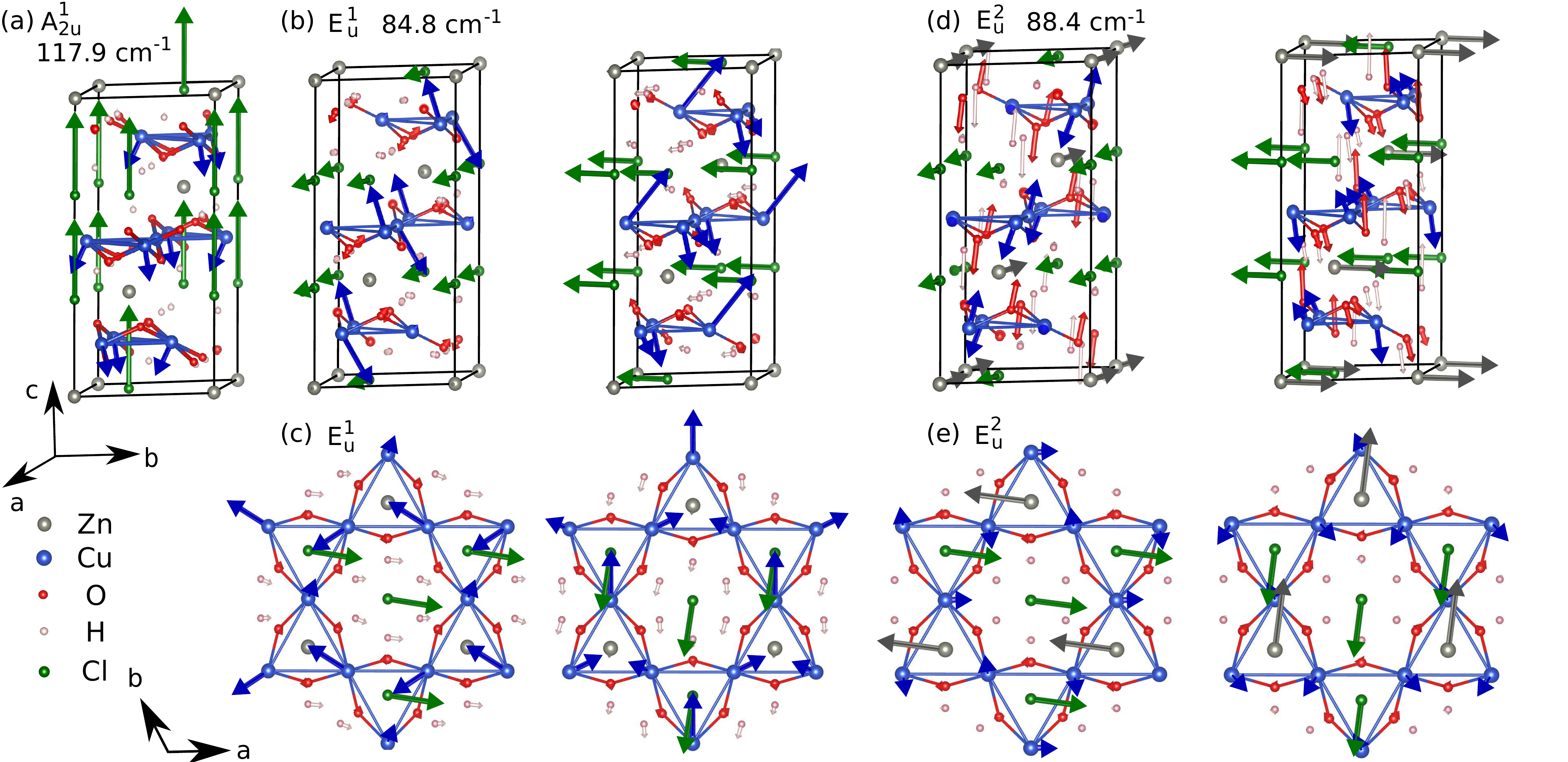}
\caption{The lowest-energy infrared vibrations according to our $ab~initio$ calculations.
	(a)  $A^1_{2u}$ mode,
	(b, d) $E_u^1$ and $E_u^2$ phonon modes and (c, e) their projection onto the $ab$-plane, respectively.}
\label{fig:move}
\end{figure*}

While qualitatively we can identify and characterize
all infrared-active phonon modes, a quantitative comparison between 
measured and calculated phonon mode frequencies
reveals some discrepancies
that we attibute to different sources. 
For instance, the measured in-plane $E^1_u$ and $E^2_u$ modes at 90 and
115~\cm\ are observed at higher frequencies than the calculated doublet, 84.8 and 88.4~\cm, respectively.
Actually, at low energies $h\nu\ll J \approx 180-190$~K, the magnetoelastic coupling
is expected to manifest most strongly, so that this discrepancy may be largely attributed to the
involvement of the spin degrees of freedom.
The mismatch is generally more pronounced
for vibrations that include deformations of the Cu-O-Cu
bond lengths and angles within the kagome layer. For both polarizations the
effect is largest for phonon modes close to 120~\cm\, which are also the modes
exhibiting anomalous temperature dependence (see Fig.~\ref{Zn_low-energy-sketch}
and Ref.~\onlinecite{Sushkov2017}). 
In contrast, the measured modes around 700--1000~\cm\ appear at lower energies than computed. The range where the phonons oscillate at higher resonance frequencies than in the calculation (0--700~\cm) coincides with the two-magnon band seen by Raman spectroscopy~\cite{Wulferding2010}.
Interestingly, also
 the high-frequency proton modes above 3000~\cm\ -- here considering the
main peak -- are subject to hardening by 5\% despite $h\nu\gg J$. 
This is attributed to the description of hydrogen in the DFT calculations.

In the following
we focus our main discussion on
those features that exhibit anomalous temperature dependence, i.e. deviations
from the common narrowing and blue-shift upon lowering the temperature.

In Fig.~\ref{Zn_low-energy-sketch} we display the three measured lowest-energy infrared-active
phonon modes of
herbertsmithite and illustrate in Fig.~\ref{fig:move} the corresponding atomic motions resulting from our calculations. The $A_{2u}$ modes are dominated by  Cl and
Cu displacements along the $c$-direction [Fig.~\ref{fig:move} (a)]. 
Panels  (b-e) show the $E_u^1$ and $E_u^2$ modes and their  $E\parallel ab$ projections 
corresponding to the measured 90 and 115~\cm\ peaks in Fig.~\ref{Zn_low-energy-sketch},
respectively. 

We detect a clear relation between the amplitude of Cu atom displacements and the
observed non-thermal broadening and red-shift which we associate
with spin-phonon coupling. 
Distortions of the kagome lattice  and, in particular,
of the CuO$_4$Cl$_2$ octahedra due to atomic displacements 
directly influence  the nearest-neighbor (super)exchange.
Since for these phonon modes the frequency scales  are
comparable to the magnetic interaction scales, it is expected that the spins sense a
modulation (distortion) of the other Cu sites which reflects in a transient
change in $J$. 
If the phonon brings about a more favourable spin configuration
lowering the total energy, the system tends to stay "longer" in this
arrangement which effectively reduces the spring constant $k$ between the
lattice sites and, thus, the resonance frequency.
In this context, we would like to discuss recent reports of a monoclinic distortion~\cite{Laurita2019,Zorko2017}.
The lowest-energy in-plane modes in
Fig.~\ref{fig:move}(b-e) indeed show Cl (and Zn) ions moving strictly within
the $ab$-direction while, at the same time, the Cu atoms are displaced in an
alternating fashion perpendicular to the kagome layers. The resulting
deformations indeed break symmetry similar to the report in
Ref.~\onlinecite{Laurita2019}. The proposed stripe phase~\cite{Laurita2019}
cannot be obtained by the in-plane projection of $E^1_u$ and $E^2_u$ but by their out-of-plane component, which
is close to the movements in the right panels of Fig.~\ref{fig:move} (b,d).
Our analysis of the phonon spectrum and atomic displacements
suggests that such measurements are compatible with $E_u$ modes lowering the symmetry to $m$ or 2, and hint to a strong magnetoelastic coupling affecting and influencing the underlying spin system. 

Further, at higher frequencies, although $h\nu\gg J$, the
proton vibrations exhibit a pronounced non-thermal behavior, too. Despite being
faster than the low-energy modes, the hydrogen atoms are bound to the oxygen
sites, which keep moving on the slow time scales of the kagome-layer phonons. Thus,
several proton oscillation cycles will be superimposed on the oxygen motions, such
as the above mentioned CuO scissoring and stretching vibrations that are
affected by the low-frequency spin-excitation continuum. This way, we expect
that also the
protons are affected by the effects of magneto-elastic coupling -- at an energy $h\nu$ an order of magnitude
larger than $J$. This feature could possibly explain the higher resonance frequency observed in experiment as compared to the computed peak position.


In summary, we performed a comprehensive theoretical and experimental study of the lattice vibrations in \Herb, providing insight into the mechanism of magneto-elastic coupling. Our $ab~initio$ calculations reproduce the vibrational peaks found in the infrared spectrum~\cite{Sushkov2017}. Quantitative differences between observed and computed resonance frequencies -- the modes seen in experiment are blue-shifted compared to the calculation -- coincide with the magnetic excitation background~\cite{Wulferding2010}. The resulting atomic displacements and oscillation amplitudes reveal an intricate relation of kagome layer deformations with anomalous broadening and red shifts of the lowest-frequency phonons upon cooling. Surprisingly, even the O--H vibrations above 3000~\cm\ 
are susceptible to the low-energy magnetic degrees of freedom -- an order of magnitude higher in frequency than $J\approx 180$~K.
Moreover, the splitting of particular phonon modes in the experimental spectrum
is attributed to the lower symmetry of the crystal resulting from disorder, providing a
handle on antisite Cu/Zn exchange in \Herb. 

We expect that optical experiments as a function of pressure and magnetic field~\cite{Kozlenko2012,Biesner2020}, as well as ultrasound studies~\cite{Wosnitza2016} may provide additional information on the nature of magneto-elastic coupling and symmetry breaking in herbertsmithite. Further, having the full information about optical and acoustic phonons allows, in principle, to properly subtract these lattice contributions from the specific heat~\cite{Helton2007} and isolate the spin entropy of the QSL.

\acknowledgements
We thank M. R. Norman for fruitful discussions. The project was supported by the Deutsche Forschungsgemeinschaft (DFG)
through grant VA117/15-1.

*Y.L. and A.P. contributed equally to this work.

\bibliography{Literature}

\begin{thebibliography}{45}%
\makeatletter
\providecommand \@ifxundefined [1]{%
 \@ifx{#1\undefined}
}%
\providecommand \@ifnum [1]{%
 \ifnum #1\expandafter \@firstoftwo
 \else \expandafter \@secondoftwo
 \fi
}%
\providecommand \@ifx [1]{%
 \ifx #1\expandafter \@firstoftwo
 \else \expandafter \@secondoftwo
 \fi
}%
\providecommand \natexlab [1]{#1}%
\providecommand \enquote  [1]{``#1''}%
\providecommand \bibnamefont  [1]{#1}%
\providecommand \bibfnamefont [1]{#1}%
\providecommand \citenamefont [1]{#1}%
\providecommand \href@noop [0]{\@secondoftwo}%
\providecommand \href [0]{\begingroup \@sanitize@url \@href}%
\providecommand \@href[1]{\@@startlink{#1}\@@href}%
\providecommand \@@href[1]{\endgroup#1\@@endlink}%
\providecommand \@sanitize@url [0]{\catcode `\\12\catcode `\$12\catcode
  `\&12\catcode `\#12\catcode `\^12\catcode `\_12\catcode `\%12\relax}%
\providecommand \@@startlink[1]{}%
\providecommand \@@endlink[0]{}%
\providecommand \url  [0]{\begingroup\@sanitize@url \@url }%
\providecommand \@url [1]{\endgroup\@href {#1}{\urlprefix }}%
\providecommand \urlprefix  [0]{URL }%
\providecommand \Eprint [0]{\href }%
\providecommand \doibase [0]{http://dx.doi.org/}%
\providecommand \selectlanguage [0]{\@gobble}%
\providecommand \bibinfo  [0]{\@secondoftwo}%
\providecommand \bibfield  [0]{\@secondoftwo}%
\providecommand \translation [1]{[#1]}%
\providecommand \BibitemOpen [0]{}%
\providecommand \bibitemStop [0]{}%
\providecommand \bibitemNoStop [0]{.\EOS\space}%
\providecommand \EOS [0]{\spacefactor3000\relax}%
\providecommand \BibitemShut  [1]{\csname bibitem#1\endcsname}%
\let\auto@bib@innerbib\@empty
\bibitem [{\citenamefont {Norman}(2016)}]{Norman2016}%
  \BibitemOpen
  \bibfield  {author} {\bibinfo {author} {\bibfnamefont {M.~R.}\ \bibnamefont
  {Norman}},\ }\href {\doibase 10.1103/RevModPhys.88.041002} {\bibfield
  {journal} {\bibinfo  {journal} {Rev. Mod. Phys.}\ }\textbf {\bibinfo {volume}
  {88}},\ \bibinfo {pages} {041002} (\bibinfo {year} {2016})}\BibitemShut
  {NoStop}%
\bibitem [{\citenamefont {Shaginyan}\ \emph {et~al.}(2019)\citenamefont
  {Shaginyan}, \citenamefont {Msezane}, \citenamefont {Amusia}, \citenamefont
  {Clark}, \citenamefont {Japaridze}, \citenamefont {Stephanovich},\ and\
  \citenamefont {Leevik}}]{Shaginyan2019}%
  \BibitemOpen
  \bibfield  {author} {\bibinfo {author} {\bibfnamefont {V.~R.}\ \bibnamefont
  {Shaginyan}}, \bibinfo {author} {\bibfnamefont {A.~Z.}\ \bibnamefont
  {Msezane}}, \bibinfo {author} {\bibfnamefont {M.~Y.}\ \bibnamefont {Amusia}},
  \bibinfo {author} {\bibfnamefont {J.~W.}\ \bibnamefont {Clark}}, \bibinfo
  {author} {\bibfnamefont {G.~S.}\ \bibnamefont {Japaridze}}, \bibinfo {author}
  {\bibfnamefont {V.~A.}\ \bibnamefont {Stephanovich}}, \ and\ \bibinfo
  {author} {\bibfnamefont {Y.~S.}\ \bibnamefont {Leevik}},\ }\href@noop {}
  {\bibfield  {journal} {\bibinfo  {journal} {Condensed Matter}\ }\textbf
  {\bibinfo {volume} {4}} (\bibinfo {year} {2019})}\BibitemShut {NoStop}%
\bibitem [{\citenamefont {Jeschke}\ \emph {et~al.}(2013)\citenamefont
  {Jeschke}, \citenamefont {Salvat-Pujol},\ and\ \citenamefont
  {Valent\'{\i}}}]{Jeschke2013}%
  \BibitemOpen
  \bibfield  {author} {\bibinfo {author} {\bibfnamefont {H.~O.}\ \bibnamefont
  {Jeschke}}, \bibinfo {author} {\bibfnamefont {F.}~\bibnamefont
  {Salvat-Pujol}}, \ and\ \bibinfo {author} {\bibfnamefont {R.}~\bibnamefont
  {Valent\'{\i}}},\ }\href {\doibase 10.1103/PhysRevB.88.075106} {\bibfield
  {journal} {\bibinfo  {journal} {Phys. Rev. B}\ }\textbf {\bibinfo {volume}
  {88}},\ \bibinfo {pages} {075106} (\bibinfo {year} {2013})}\BibitemShut
  {NoStop}%
\bibitem [{\citenamefont {Mendels}\ \emph {et~al.}(2007)\citenamefont
  {Mendels}, \citenamefont {Bert}, \citenamefont {de~Vries}, \citenamefont
  {Olariu}, \citenamefont {Harrison}, \citenamefont {Duc}, \citenamefont
  {Trombe}, \citenamefont {Lord}, \citenamefont {Amato},\ and\ \citenamefont
  {Baines}}]{Mendels2007}%
  \BibitemOpen
  \bibfield  {author} {\bibinfo {author} {\bibfnamefont {P.}~\bibnamefont
  {Mendels}}, \bibinfo {author} {\bibfnamefont {F.}~\bibnamefont {Bert}},
  \bibinfo {author} {\bibfnamefont {M.~A.}\ \bibnamefont {de~Vries}}, \bibinfo
  {author} {\bibfnamefont {A.}~\bibnamefont {Olariu}}, \bibinfo {author}
  {\bibfnamefont {A.}~\bibnamefont {Harrison}}, \bibinfo {author}
  {\bibfnamefont {F.}~\bibnamefont {Duc}}, \bibinfo {author} {\bibfnamefont
  {J.~C.}\ \bibnamefont {Trombe}}, \bibinfo {author} {\bibfnamefont {J.~S.}\
  \bibnamefont {Lord}}, \bibinfo {author} {\bibfnamefont {A.}~\bibnamefont
  {Amato}}, \ and\ \bibinfo {author} {\bibfnamefont {C.}~\bibnamefont
  {Baines}},\ }\href {\doibase 10.1103/PhysRevLett.98.077204} {\bibfield
  {journal} {\bibinfo  {journal} {Phys. Rev. Lett.}\ }\textbf {\bibinfo
  {volume} {98}},\ \bibinfo {pages} {077204} (\bibinfo {year}
  {2007})}\BibitemShut {NoStop}%
\bibitem [{\citenamefont {Han}\ \emph {et~al.}(2012)\citenamefont {Han},
  \citenamefont {Helton}, \citenamefont {Chu}, \citenamefont {Nocera},
  \citenamefont {Rodriguez-Rivera}, \citenamefont {Broholm},\ and\
  \citenamefont {Lee}}]{Han2012}%
  \BibitemOpen
  \bibfield  {author} {\bibinfo {author} {\bibfnamefont {T.-H.}\ \bibnamefont
  {Han}}, \bibinfo {author} {\bibfnamefont {J.~S.}\ \bibnamefont {Helton}},
  \bibinfo {author} {\bibfnamefont {S.}~\bibnamefont {Chu}}, \bibinfo {author}
  {\bibfnamefont {D.~G.}\ \bibnamefont {Nocera}}, \bibinfo {author}
  {\bibfnamefont {J.~A.}\ \bibnamefont {Rodriguez-Rivera}}, \bibinfo {author}
  {\bibfnamefont {C.}~\bibnamefont {Broholm}}, \ and\ \bibinfo {author}
  {\bibfnamefont {Y.~S.}\ \bibnamefont {Lee}},\ }\href {\doibase
  10.1038/nature11659} {\bibfield  {journal} {\bibinfo  {journal} {Nature}\
  }\textbf {\bibinfo {volume} {492}},\ \bibinfo {pages} {406} (\bibinfo {year}
  {2012})}\BibitemShut {NoStop}%
\bibitem [{\citenamefont {Shores}\ \emph {et~al.}(2005)\citenamefont {Shores},
  \citenamefont {Nytko}, \citenamefont {Bartlett},\ and\ \citenamefont
  {Nocera}}]{Shores2005}%
  \BibitemOpen
  \bibfield  {author} {\bibinfo {author} {\bibfnamefont {M.~P.}\ \bibnamefont
  {Shores}}, \bibinfo {author} {\bibfnamefont {E.~A.}\ \bibnamefont {Nytko}},
  \bibinfo {author} {\bibfnamefont {B.~M.}\ \bibnamefont {Bartlett}}, \ and\
  \bibinfo {author} {\bibfnamefont {D.~G.}\ \bibnamefont {Nocera}},\ }\href
  {\doibase 10.1021/ja053891p} {\bibfield  {journal} {\bibinfo  {journal} {J.
  Am. Chem. Soc.}\ }\textbf {\bibinfo {volume} {127}},\ \bibinfo {pages}
  {13462} (\bibinfo {year} {2005})}\BibitemShut {NoStop}%
\bibitem [{\citenamefont {Helton}\ \emph {et~al.}(2007)\citenamefont {Helton},
  \citenamefont {Matan}, \citenamefont {Shores}, \citenamefont {Nytko},
  \citenamefont {Bartlett}, \citenamefont {Yoshida}, \citenamefont {Takano},
  \citenamefont {Suslov}, \citenamefont {Qiu}, \citenamefont {Chung},
  \citenamefont {Nocera},\ and\ \citenamefont {Lee}}]{Helton2007}%
  \BibitemOpen
  \bibfield  {author} {\bibinfo {author} {\bibfnamefont {J.~S.}\ \bibnamefont
  {Helton}}, \bibinfo {author} {\bibfnamefont {K.}~\bibnamefont {Matan}},
  \bibinfo {author} {\bibfnamefont {M.~P.}\ \bibnamefont {Shores}}, \bibinfo
  {author} {\bibfnamefont {E.~A.}\ \bibnamefont {Nytko}}, \bibinfo {author}
  {\bibfnamefont {B.~M.}\ \bibnamefont {Bartlett}}, \bibinfo {author}
  {\bibfnamefont {Y.}~\bibnamefont {Yoshida}}, \bibinfo {author} {\bibfnamefont
  {Y.}~\bibnamefont {Takano}}, \bibinfo {author} {\bibfnamefont
  {A.}~\bibnamefont {Suslov}}, \bibinfo {author} {\bibfnamefont
  {Y.}~\bibnamefont {Qiu}}, \bibinfo {author} {\bibfnamefont {J.-H.}\
  \bibnamefont {Chung}}, \bibinfo {author} {\bibfnamefont {D.~G.}\ \bibnamefont
  {Nocera}}, \ and\ \bibinfo {author} {\bibfnamefont {Y.~S.}\ \bibnamefont
  {Lee}},\ }\href {https://link.aps.org/doi/10.1103/PhysRevLett.98.107204}
  {\bibfield  {journal} {\bibinfo  {journal} {Phys. Rev. Lett.}\ }\textbf
  {\bibinfo {volume} {98}},\ \bibinfo {pages} {107204} (\bibinfo {year}
  {2007})}\BibitemShut {NoStop}%
\bibitem [{\citenamefont {de~Vries}\ \emph {et~al.}(2009)\citenamefont
  {de~Vries}, \citenamefont {Stewart}, \citenamefont {Deen}, \citenamefont
  {Piatek}, \citenamefont {Nilsen}, \citenamefont {R{\o}nnow},\ and\
  \citenamefont {Harrison}}]{DeVries2009}%
  \BibitemOpen
  \bibfield  {author} {\bibinfo {author} {\bibfnamefont {M.~A.}\ \bibnamefont
  {de~Vries}}, \bibinfo {author} {\bibfnamefont {J.~R.}\ \bibnamefont
  {Stewart}}, \bibinfo {author} {\bibfnamefont {P.~P.}\ \bibnamefont {Deen}},
  \bibinfo {author} {\bibfnamefont {J.~O.}\ \bibnamefont {Piatek}}, \bibinfo
  {author} {\bibfnamefont {G.~J.}\ \bibnamefont {Nilsen}}, \bibinfo {author}
  {\bibfnamefont {H.~M.}\ \bibnamefont {R{\o}nnow}}, \ and\ \bibinfo {author}
  {\bibfnamefont {A.}~\bibnamefont {Harrison}},\ }\href
  {https://link.aps.org/doi/10.1103/PhysRevLett.103.237201} {\bibfield
  {journal} {\bibinfo  {journal} {Phys. Rev. Lett.}\ }\textbf {\bibinfo
  {volume} {103}},\ \bibinfo {pages} {237201} (\bibinfo {year}
  {2009})}\BibitemShut {NoStop}%
\bibitem [{\citenamefont {Imai}\ \emph {et~al.}(2008)\citenamefont {Imai},
  \citenamefont {Nytko}, \citenamefont {Bartlett}, \citenamefont {Shores},\
  and\ \citenamefont {Nocera}}]{Imai2008}%
  \BibitemOpen
  \bibfield  {author} {\bibinfo {author} {\bibfnamefont {T.}~\bibnamefont
  {Imai}}, \bibinfo {author} {\bibfnamefont {E.~A.}\ \bibnamefont {Nytko}},
  \bibinfo {author} {\bibfnamefont {B.~M.}\ \bibnamefont {Bartlett}}, \bibinfo
  {author} {\bibfnamefont {M.~P.}\ \bibnamefont {Shores}}, \ and\ \bibinfo
  {author} {\bibfnamefont {D.~G.}\ \bibnamefont {Nocera}},\ }\href {\doibase
  10.1103/PhysRevLett.100.077203} {\bibfield  {journal} {\bibinfo  {journal}
  {Phys. Rev. Lett.}\ }\textbf {\bibinfo {volume} {100}},\ \bibinfo {pages}
  {77203} (\bibinfo {year} {2008})}\BibitemShut {NoStop}%
\bibitem [{\citenamefont {Olariu}\ \emph {et~al.}(2008)\citenamefont {Olariu},
  \citenamefont {Mendels}, \citenamefont {Bert}, \citenamefont {Duc},
  \citenamefont {Trombe}, \citenamefont {de~Vries},\ and\ \citenamefont
  {Harrison}}]{Olariu2008}%
  \BibitemOpen
  \bibfield  {author} {\bibinfo {author} {\bibfnamefont {A.}~\bibnamefont
  {Olariu}}, \bibinfo {author} {\bibfnamefont {P.}~\bibnamefont {Mendels}},
  \bibinfo {author} {\bibfnamefont {F.}~\bibnamefont {Bert}}, \bibinfo {author}
  {\bibfnamefont {F.}~\bibnamefont {Duc}}, \bibinfo {author} {\bibfnamefont
  {J.~C.}\ \bibnamefont {Trombe}}, \bibinfo {author} {\bibfnamefont {M.~A.}\
  \bibnamefont {de~Vries}}, \ and\ \bibinfo {author} {\bibfnamefont
  {A.}~\bibnamefont {Harrison}},\ }\href
  {https://link.aps.org/doi/10.1103/PhysRevLett.100.087202} {\bibfield
  {journal} {\bibinfo  {journal} {Phys. Rev. Lett.}\ }\textbf {\bibinfo
  {volume} {100}},\ \bibinfo {pages} {87202} (\bibinfo {year}
  {2008})}\BibitemShut {NoStop}%
\bibitem [{\citenamefont {Singh}\ and\ \citenamefont {Huse}(2007)}]{Singh2007}%
  \BibitemOpen
  \bibfield  {author} {\bibinfo {author} {\bibfnamefont {R.~R.~P.}\
  \bibnamefont {Singh}}\ and\ \bibinfo {author} {\bibfnamefont {D.~A.}\
  \bibnamefont {Huse}},\ }\href {\doibase 10.1103/PhysRevB.76.180407}
  {\bibfield  {journal} {\bibinfo  {journal} {Phys. Rev. B}\ }\textbf {\bibinfo
  {volume} {76}},\ \bibinfo {pages} {180407} (\bibinfo {year}
  {2007})}\BibitemShut {NoStop}%
\bibitem [{\citenamefont {Ran}\ \emph {et~al.}(2007)\citenamefont {Ran},
  \citenamefont {Hermele}, \citenamefont {Lee},\ and\ \citenamefont
  {Wen}}]{Ran2007}%
  \BibitemOpen
  \bibfield  {author} {\bibinfo {author} {\bibfnamefont {Y.}~\bibnamefont
  {Ran}}, \bibinfo {author} {\bibfnamefont {M.}~\bibnamefont {Hermele}},
  \bibinfo {author} {\bibfnamefont {P.~A.}\ \bibnamefont {Lee}}, \ and\
  \bibinfo {author} {\bibfnamefont {X.-G.}\ \bibnamefont {Wen}},\ }\href
  {\doibase 10.1103/PhysRevLett.98.117205} {\bibfield  {journal} {\bibinfo
  {journal} {Phys. Rev. Lett.}\ }\textbf {\bibinfo {volume} {98}},\ \bibinfo
  {pages} {117205} (\bibinfo {year} {2007})}\BibitemShut {NoStop}%
\bibitem [{\citenamefont {Depenbrock}\ \emph {et~al.}(2012)\citenamefont
  {Depenbrock}, \citenamefont {McCulloch},\ and\ \citenamefont
  {Schollw\"ock}}]{Depenbrock2012}%
  \BibitemOpen
  \bibfield  {author} {\bibinfo {author} {\bibfnamefont {S.}~\bibnamefont
  {Depenbrock}}, \bibinfo {author} {\bibfnamefont {I.~P.}\ \bibnamefont
  {McCulloch}}, \ and\ \bibinfo {author} {\bibfnamefont {U.}~\bibnamefont
  {Schollw\"ock}},\ }\href {\doibase 10.1103/PhysRevLett.109.067201} {\bibfield
   {journal} {\bibinfo  {journal} {Phys. Rev. Lett.}\ }\textbf {\bibinfo
  {volume} {109}},\ \bibinfo {pages} {067201} (\bibinfo {year}
  {2012})}\BibitemShut {NoStop}%
\bibitem [{\citenamefont {Iqbal}\ \emph {et~al.}(2013)\citenamefont {Iqbal},
  \citenamefont {Becca}, \citenamefont {Sorella},\ and\ \citenamefont
  {Poilblanc}}]{Iqbal2013}%
  \BibitemOpen
  \bibfield  {author} {\bibinfo {author} {\bibfnamefont {Y.}~\bibnamefont
  {Iqbal}}, \bibinfo {author} {\bibfnamefont {F.}~\bibnamefont {Becca}},
  \bibinfo {author} {\bibfnamefont {S.}~\bibnamefont {Sorella}}, \ and\
  \bibinfo {author} {\bibfnamefont {D.}~\bibnamefont {Poilblanc}},\ }\href
  {\doibase 10.1103/PhysRevB.87.060405} {\bibfield  {journal} {\bibinfo
  {journal} {Phys. Rev. B}\ }\textbf {\bibinfo {volume} {87}},\ \bibinfo
  {pages} {060405} (\bibinfo {year} {2013})}\BibitemShut {NoStop}%
\bibitem [{\citenamefont {Fu}\ \emph {et~al.}(2015)\citenamefont {Fu},
  \citenamefont {Imai}, \citenamefont {Han},\ and\ \citenamefont
  {Lee}}]{Fu2015}%
  \BibitemOpen
  \bibfield  {author} {\bibinfo {author} {\bibfnamefont {M.}~\bibnamefont
  {Fu}}, \bibinfo {author} {\bibfnamefont {T.}~\bibnamefont {Imai}}, \bibinfo
  {author} {\bibfnamefont {T.-H.}\ \bibnamefont {Han}}, \ and\ \bibinfo
  {author} {\bibfnamefont {Y.~S.}\ \bibnamefont {Lee}},\ }\href
  {http://science.sciencemag.org/content/350/6261/655.abstract} {\bibfield
  {journal} {\bibinfo  {journal} {Science}\ }\textbf {\bibinfo {volume}
  {350}},\ \bibinfo {pages} {655 LP } (\bibinfo {year} {2015})}\BibitemShut
  {NoStop}%
\bibitem [{\citenamefont {He}\ \emph {et~al.}(2017)\citenamefont {He},
  \citenamefont {Zaletel}, \citenamefont {Oshikawa},\ and\ \citenamefont
  {Pollmann}}]{He2017}%
  \BibitemOpen
  \bibfield  {author} {\bibinfo {author} {\bibfnamefont {Y.-C.}\ \bibnamefont
  {He}}, \bibinfo {author} {\bibfnamefont {M.~P.}\ \bibnamefont {Zaletel}},
  \bibinfo {author} {\bibfnamefont {M.}~\bibnamefont {Oshikawa}}, \ and\
  \bibinfo {author} {\bibfnamefont {F.}~\bibnamefont {Pollmann}},\ }\href
  {\doibase 10.1103/PhysRevX.7.031020} {\bibfield  {journal} {\bibinfo
  {journal} {Phys. Rev. X}\ }\textbf {\bibinfo {volume} {7}},\ \bibinfo {pages}
  {031020} (\bibinfo {year} {2017})}\BibitemShut {NoStop}%
\bibitem [{\citenamefont {Liao}\ \emph {et~al.}(2017)\citenamefont {Liao},
  \citenamefont {Xie}, \citenamefont {Chen}, \citenamefont {Liu}, \citenamefont
  {Xie}, \citenamefont {Huang}, \citenamefont {Normand},\ and\ \citenamefont
  {Xiang}}]{Liao2017}%
  \BibitemOpen
  \bibfield  {author} {\bibinfo {author} {\bibfnamefont {H.~J.}\ \bibnamefont
  {Liao}}, \bibinfo {author} {\bibfnamefont {Z.~Y.}\ \bibnamefont {Xie}},
  \bibinfo {author} {\bibfnamefont {J.}~\bibnamefont {Chen}}, \bibinfo {author}
  {\bibfnamefont {Z.~Y.}\ \bibnamefont {Liu}}, \bibinfo {author} {\bibfnamefont
  {H.~D.}\ \bibnamefont {Xie}}, \bibinfo {author} {\bibfnamefont {R.~Z.}\
  \bibnamefont {Huang}}, \bibinfo {author} {\bibfnamefont {B.}~\bibnamefont
  {Normand}}, \ and\ \bibinfo {author} {\bibfnamefont {T.}~\bibnamefont
  {Xiang}},\ }\href {\doibase 10.1103/PhysRevLett.118.137202} {\bibfield
  {journal} {\bibinfo  {journal} {Phys. Rev. Lett.}\ }\textbf {\bibinfo
  {volume} {118}},\ \bibinfo {pages} {137202} (\bibinfo {year}
  {2017})}\BibitemShut {NoStop}%
\bibitem [{\citenamefont {Khuntia}\ \emph {et~al.}(2019)\citenamefont
  {Khuntia}, \citenamefont {Velazquez}, \citenamefont {Barthélemy},
  \citenamefont {Bert}, \citenamefont {Kermarrec}, \citenamefont {Legros},
  \citenamefont {Bernu}, \citenamefont {Messio}, \citenamefont {Zorko},\ and\
  \citenamefont {Mendels}}]{Khuntia2019}%
  \BibitemOpen
  \bibfield  {author} {\bibinfo {author} {\bibfnamefont {P.}~\bibnamefont
  {Khuntia}}, \bibinfo {author} {\bibfnamefont {M.}~\bibnamefont {Velazquez}},
  \bibinfo {author} {\bibfnamefont {Q.}~\bibnamefont {Barthélemy}}, \bibinfo
  {author} {\bibfnamefont {F.}~\bibnamefont {Bert}}, \bibinfo {author}
  {\bibfnamefont {E.}~\bibnamefont {Kermarrec}}, \bibinfo {author}
  {\bibfnamefont {A.}~\bibnamefont {Legros}}, \bibinfo {author} {\bibfnamefont
  {B.}~\bibnamefont {Bernu}}, \bibinfo {author} {\bibfnamefont
  {L.}~\bibnamefont {Messio}}, \bibinfo {author} {\bibfnamefont
  {A.}~\bibnamefont {Zorko}}, \ and\ \bibinfo {author} {\bibfnamefont
  {P.}~\bibnamefont {Mendels}},\ }\href@noop {} {\bibfield  {journal} {\bibinfo
   {journal} {arXiv:1911.09552}\ } (\bibinfo {year} {2019})}\BibitemShut
  {NoStop}%
\bibitem [{\citenamefont {Laurita}\ \emph {et~al.}(2019)\citenamefont
  {Laurita}, \citenamefont {Ron}, \citenamefont {Han}, \citenamefont {Scheie},
  \citenamefont {Sheckelton}, \citenamefont {Smaha}, , \citenamefont {He},
  \citenamefont {Wen}, \citenamefont {Lee}, \citenamefont {Lee}, \citenamefont
  {Norman},\ and\ \citenamefont {Hsieh}}]{Laurita2019}%
  \BibitemOpen
  \bibfield  {author} {\bibinfo {author} {\bibfnamefont {N.~J.}\ \bibnamefont
  {Laurita}}, \bibinfo {author} {\bibfnamefont {A.}~\bibnamefont {Ron}},
  \bibinfo {author} {\bibfnamefont {J.~W.}\ \bibnamefont {Han}}, \bibinfo
  {author} {\bibfnamefont {A.}~\bibnamefont {Scheie}}, \bibinfo {author}
  {\bibfnamefont {J.~P.}\ \bibnamefont {Sheckelton}}, \bibinfo {author}
  {\bibfnamefont {R.~W.}\ \bibnamefont {Smaha}}, , \bibinfo {author}
  {\bibfnamefont {W.}~\bibnamefont {He}}, \bibinfo {author} {\bibfnamefont
  {J.-J.}\ \bibnamefont {Wen}}, \bibinfo {author} {\bibfnamefont {J.~S.}\
  \bibnamefont {Lee}}, \bibinfo {author} {\bibfnamefont {Y.~S.}\ \bibnamefont
  {Lee}}, \bibinfo {author} {\bibfnamefont {M.~R.}\ \bibnamefont {Norman}}, \
  and\ \bibinfo {author} {\bibfnamefont {D.}~\bibnamefont {Hsieh}},\
  }\href@noop {} {\bibfield  {journal} {\bibinfo  {journal} {arXiv:1910.13606}\
  } (\bibinfo {year} {2019})}\BibitemShut {NoStop}%
\bibitem [{\citenamefont {Zorko}\ \emph {et~al.}(2017)\citenamefont {Zorko},
  \citenamefont {Herak}, \citenamefont {Gomil{\v{s}}ek}, \citenamefont {van
  Tol}, \citenamefont {Vel{\'{a}}zquez}, \citenamefont {Khuntia}, \citenamefont
  {Bert},\ and\ \citenamefont {Mendels}}]{Zorko2017}%
  \BibitemOpen
  \bibfield  {author} {\bibinfo {author} {\bibfnamefont {A.}~\bibnamefont
  {Zorko}}, \bibinfo {author} {\bibfnamefont {M.}~\bibnamefont {Herak}},
  \bibinfo {author} {\bibfnamefont {M.}~\bibnamefont {Gomil{\v{s}}ek}},
  \bibinfo {author} {\bibfnamefont {J.}~\bibnamefont {van Tol}}, \bibinfo
  {author} {\bibfnamefont {M.}~\bibnamefont {Vel{\'{a}}zquez}}, \bibinfo
  {author} {\bibfnamefont {P.}~\bibnamefont {Khuntia}}, \bibinfo {author}
  {\bibfnamefont {F.}~\bibnamefont {Bert}}, \ and\ \bibinfo {author}
  {\bibfnamefont {P.}~\bibnamefont {Mendels}},\ }\href
  {https://link.aps.org/doi/10.1103/PhysRevLett.118.017202} {\bibfield
  {journal} {\bibinfo  {journal} {Phys. Rev. Lett.}\ }\textbf {\bibinfo
  {volume} {118}},\ \bibinfo {pages} {17202} (\bibinfo {year}
  {2017})}\BibitemShut {NoStop}%
\bibitem [{\citenamefont {Sushkov}\ \emph {et~al.}(2017)\citenamefont
  {Sushkov}, \citenamefont {Jenkins}, \citenamefont {Han}, \citenamefont
  {Lee},\ and\ \citenamefont {D}}]{Sushkov2017}%
  \BibitemOpen
  \bibfield  {author} {\bibinfo {author} {\bibfnamefont {A.~B.}\ \bibnamefont
  {Sushkov}}, \bibinfo {author} {\bibfnamefont {G.~S.}\ \bibnamefont
  {Jenkins}}, \bibinfo {author} {\bibfnamefont {T.-H.}\ \bibnamefont {Han}},
  \bibinfo {author} {\bibfnamefont {Y.~S.}\ \bibnamefont {Lee}}, \ and\
  \bibinfo {author} {\bibfnamefont {D.~H.}\ \bibnamefont {D}},\ }\href
  {http://stacks.iop.org/0953-8984/29/i=9/a=095802} {\bibfield  {journal}
  {\bibinfo  {journal} {J. Phys. Condens. Matter}\ }\textbf {\bibinfo {volume}
  {29}},\ \bibinfo {pages} {95802} (\bibinfo {year} {2017})}\BibitemShut
  {NoStop}%
\bibitem [{\citenamefont {Pustogow}\ \emph {et~al.}(2017)\citenamefont
  {Pustogow}, \citenamefont {Li}, \citenamefont {Voloshenko}, \citenamefont
  {Puphal}, \citenamefont {Krellner}, \citenamefont {Mazin}, \citenamefont
  {Dressel},\ and\ \citenamefont {Valent\'i}}]{Pustogow2017Herbertsmithite}%
  \BibitemOpen
  \bibfield  {author} {\bibinfo {author} {\bibfnamefont {A.}~\bibnamefont
  {Pustogow}}, \bibinfo {author} {\bibfnamefont {Y.}~\bibnamefont {Li}},
  \bibinfo {author} {\bibfnamefont {I.}~\bibnamefont {Voloshenko}}, \bibinfo
  {author} {\bibfnamefont {P.}~\bibnamefont {Puphal}}, \bibinfo {author}
  {\bibfnamefont {C.}~\bibnamefont {Krellner}}, \bibinfo {author}
  {\bibfnamefont {I.~I.}\ \bibnamefont {Mazin}}, \bibinfo {author}
  {\bibfnamefont {M.}~\bibnamefont {Dressel}}, \ and\ \bibinfo {author}
  {\bibfnamefont {R.}~\bibnamefont {Valent\'i}},\ }\href@noop {} {\bibfield
  {journal} {\bibinfo  {journal} {Phys. Rev. B}\ }\textbf {\bibinfo {volume}
  {96}},\ \bibinfo {pages} {241114(R)} (\bibinfo {year} {2017})}\BibitemShut
  {NoStop}%
\bibitem [{\citenamefont {Wulferding}\ \emph {et~al.}(2010)\citenamefont
  {Wulferding}, \citenamefont {Lemmens}, \citenamefont {Scheib}, \citenamefont
  {R{\"{o}}der}, \citenamefont {Mendels}, \citenamefont {Chu}, \citenamefont
  {Han},\ and\ \citenamefont {Lee}}]{Wulferding2010}%
  \BibitemOpen
  \bibfield  {author} {\bibinfo {author} {\bibfnamefont {D.}~\bibnamefont
  {Wulferding}}, \bibinfo {author} {\bibfnamefont {P.}~\bibnamefont {Lemmens}},
  \bibinfo {author} {\bibfnamefont {P.}~\bibnamefont {Scheib}}, \bibinfo
  {author} {\bibfnamefont {J.}~\bibnamefont {R{\"{o}}der}}, \bibinfo {author}
  {\bibfnamefont {P.}~\bibnamefont {Mendels}}, \bibinfo {author} {\bibfnamefont
  {S.}~\bibnamefont {Chu}}, \bibinfo {author} {\bibfnamefont {T.}~\bibnamefont
  {Han}}, \ and\ \bibinfo {author} {\bibfnamefont {Y.~S.}\ \bibnamefont
  {Lee}},\ }\href {https://link.aps.org/doi/10.1103/PhysRevB.82.144412}
  {\bibfield  {journal} {\bibinfo  {journal} {Phys. Rev. B}\ }\textbf {\bibinfo
  {volume} {82}},\ \bibinfo {pages} {144412} (\bibinfo {year}
  {2010})}\BibitemShut {NoStop}%
\bibitem [{\citenamefont {Sushkov}\ \emph {et~al.}(2005)\citenamefont
  {Sushkov}, \citenamefont {Tchernyshyov}, \citenamefont {II}, \citenamefont
  {Cheong},\ and\ \citenamefont {Drew}}]{Sushkov2005}%
  \BibitemOpen
  \bibfield  {author} {\bibinfo {author} {\bibfnamefont {A.~B.}\ \bibnamefont
  {Sushkov}}, \bibinfo {author} {\bibfnamefont {O.}~\bibnamefont
  {Tchernyshyov}}, \bibinfo {author} {\bibfnamefont {W.~R.}\ \bibnamefont
  {II}}, \bibinfo {author} {\bibfnamefont {S.~W.}\ \bibnamefont {Cheong}}, \
  and\ \bibinfo {author} {\bibfnamefont {H.~D.}\ \bibnamefont {Drew}},\ }\href
  {https://link.aps.org/doi/10.1103/PhysRevLett.94.137202} {\bibfield
  {journal} {\bibinfo  {journal} {Phys. Rev. Lett.}\ }\textbf {\bibinfo
  {volume} {94}},\ \bibinfo {pages} {137202} (\bibinfo {year}
  {2005})}\BibitemShut {NoStop}%
\bibitem [{\citenamefont {Sandilands}\ \emph {et~al.}(2015)\citenamefont
  {Sandilands}, \citenamefont {Tian}, \citenamefont {Plumb}, \citenamefont
  {Kim},\ and\ \citenamefont {Burch}}]{Sandilands2015}%
  \BibitemOpen
  \bibfield  {author} {\bibinfo {author} {\bibfnamefont {L.~J.}\ \bibnamefont
  {Sandilands}}, \bibinfo {author} {\bibfnamefont {Y.}~\bibnamefont {Tian}},
  \bibinfo {author} {\bibfnamefont {K.~W.}\ \bibnamefont {Plumb}}, \bibinfo
  {author} {\bibfnamefont {Y.-J.}\ \bibnamefont {Kim}}, \ and\ \bibinfo
  {author} {\bibfnamefont {K.~S.}\ \bibnamefont {Burch}},\ }\href
  {https://link.aps.org/doi/10.1103/PhysRevLett.114.147201} {\bibfield
  {journal} {\bibinfo  {journal} {Phys. Rev. Lett.}\ }\textbf {\bibinfo
  {volume} {114}},\ \bibinfo {pages} {147201} (\bibinfo {year}
  {2015})}\BibitemShut {NoStop}%
\bibitem [{\citenamefont {Han}\ \emph {et~al.}(2011)\citenamefont {Han},
  \citenamefont {Helton}, \citenamefont {Chu}, \citenamefont {Prodi},
  \citenamefont {Singh}, \citenamefont {Mazzoli}, \citenamefont {Mu\"uller},
  \citenamefont {Nocera},\ and\ \citenamefont {Lee}}]{Han2011}%
  \BibitemOpen
  \bibfield  {author} {\bibinfo {author} {\bibfnamefont {T.~H.}\ \bibnamefont
  {Han}}, \bibinfo {author} {\bibfnamefont {J.~S.}\ \bibnamefont {Helton}},
  \bibinfo {author} {\bibfnamefont {S.}~\bibnamefont {Chu}}, \bibinfo {author}
  {\bibfnamefont {A.}~\bibnamefont {Prodi}}, \bibinfo {author} {\bibfnamefont
  {D.~K.}\ \bibnamefont {Singh}}, \bibinfo {author} {\bibfnamefont
  {C.}~\bibnamefont {Mazzoli}}, \bibinfo {author} {\bibfnamefont
  {P.}~\bibnamefont {Mu\"uller}}, \bibinfo {author} {\bibfnamefont {D.~G.}\
  \bibnamefont {Nocera}}, \ and\ \bibinfo {author} {\bibfnamefont {Y.~S.}\
  \bibnamefont {Lee}},\ }\href {\doibase 10.1103/PhysRevB.83.100402} {\bibfield
   {journal} {\bibinfo  {journal} {Phys. Rev. B}\ }\textbf {\bibinfo {volume}
  {83}},\ \bibinfo {pages} {100402} (\bibinfo {year} {2011})}\BibitemShut
  {NoStop}%
\bibitem [{\citenamefont {Braithwaite}\ \emph {et~al.}(2004)\citenamefont
  {Braithwaite}, \citenamefont {Mereiter}, \citenamefont {Paar},\ and\
  \citenamefont {Clark}}]{Braithwaite2004}%
  \BibitemOpen
  \bibfield  {author} {\bibinfo {author} {\bibfnamefont {R.~S.~W.}\
  \bibnamefont {Braithwaite}}, \bibinfo {author} {\bibfnamefont
  {K.}~\bibnamefont {Mereiter}}, \bibinfo {author} {\bibfnamefont {W.~H.}\
  \bibnamefont {Paar}}, \ and\ \bibinfo {author} {\bibfnamefont {A.~M.}\
  \bibnamefont {Clark}},\ }\href
  {http://minmag.geoscienceworld.org/content/68/3/527/figures-only} {\bibfield
  {journal} {\bibinfo  {journal} {Mineral. Mag.}\ }\textbf {\bibinfo {volume}
  {68}} (\bibinfo {year} {2004})}\BibitemShut {NoStop}%
\bibitem [{\citenamefont {Togo}\ \emph {et~al.}(2008)\citenamefont {Togo},
  \citenamefont {Oba},\ and\ \citenamefont {Tanaka}}]{Togo2008}%
  \BibitemOpen
  \bibfield  {author} {\bibinfo {author} {\bibfnamefont {A.}~\bibnamefont
  {Togo}}, \bibinfo {author} {\bibfnamefont {F.}~\bibnamefont {Oba}}, \ and\
  \bibinfo {author} {\bibfnamefont {I.}~\bibnamefont {Tanaka}},\ }\href
  {\doibase 10.1103/PhysRevB.78.134106} {\bibfield  {journal} {\bibinfo
  {journal} {Phys. Rev. B}\ }\textbf {\bibinfo {volume} {78}},\ \bibinfo
  {pages} {134106} (\bibinfo {year} {2008})}\BibitemShut {NoStop}%
\bibitem [{\citenamefont {Togo}\ and\ \citenamefont {Tanaka}(2015)}]{Togo2015}%
  \BibitemOpen
  \bibfield  {author} {\bibinfo {author} {\bibfnamefont {A.}~\bibnamefont
  {Togo}}\ and\ \bibinfo {author} {\bibfnamefont {I.}~\bibnamefont {Tanaka}},\
  }\href {\doibase https://doi.org/10.1016/j.scriptamat.2015.07.021} {\bibfield
   {journal} {\bibinfo  {journal} {Scripta Materialia}\ }\textbf {\bibinfo
  {volume} {108}},\ \bibinfo {pages} {1 } (\bibinfo {year} {2015})}\BibitemShut
  {NoStop}%
\bibitem [{\citenamefont {Parlinski}\ \emph {et~al.}(1997)\citenamefont
  {Parlinski}, \citenamefont {Li},\ and\ \citenamefont
  {Kawazoe}}]{Parlinski1997}%
  \BibitemOpen
  \bibfield  {author} {\bibinfo {author} {\bibfnamefont {K.}~\bibnamefont
  {Parlinski}}, \bibinfo {author} {\bibfnamefont {Z.~Q.}\ \bibnamefont {Li}}, \
  and\ \bibinfo {author} {\bibfnamefont {Y.}~\bibnamefont {Kawazoe}},\ }\href
  {\doibase 10.1103/PhysRevLett.78.4063} {\bibfield  {journal} {\bibinfo
  {journal} {Phys. Rev. Lett.}\ }\textbf {\bibinfo {volume} {78}},\ \bibinfo
  {pages} {4063} (\bibinfo {year} {1997})}\BibitemShut {NoStop}%
\bibitem [{\citenamefont {Perdew}\ \emph {et~al.}(1996)\citenamefont {Perdew},
  \citenamefont {Burke},\ and\ \citenamefont {Ernzerhof}}]{Perdew1996}%
  \BibitemOpen
  \bibfield  {author} {\bibinfo {author} {\bibfnamefont {J.~P.}\ \bibnamefont
  {Perdew}}, \bibinfo {author} {\bibfnamefont {K.}~\bibnamefont {Burke}}, \
  and\ \bibinfo {author} {\bibfnamefont {M.}~\bibnamefont {Ernzerhof}},\ }\href
  {\doibase 10.1103/PhysRevLett.77.3865} {\bibfield  {journal} {\bibinfo
  {journal} {Phys. Rev. Lett.}\ }\textbf {\bibinfo {volume} {77}},\ \bibinfo
  {pages} {3865} (\bibinfo {year} {1996})}\BibitemShut {NoStop}%
\bibitem [{\citenamefont {Bl\"ochl}(1994)}]{Bloechl1994}%
  \BibitemOpen
  \bibfield  {author} {\bibinfo {author} {\bibfnamefont {P.~E.}\ \bibnamefont
  {Bl\"ochl}},\ }\href {\doibase 10.1103/PhysRevB.50.17953} {\bibfield
  {journal} {\bibinfo  {journal} {Phys. Rev. B}\ }\textbf {\bibinfo {volume}
  {50}},\ \bibinfo {pages} {17953} (\bibinfo {year} {1994})}\BibitemShut
  {NoStop}%
\bibitem [{\citenamefont {Kresse}\ and\ \citenamefont
  {Hafner}(1993)}]{Kresse1993}%
  \BibitemOpen
  \bibfield  {author} {\bibinfo {author} {\bibfnamefont {G.}~\bibnamefont
  {Kresse}}\ and\ \bibinfo {author} {\bibfnamefont {J.}~\bibnamefont
  {Hafner}},\ }\href {\doibase 10.1103/PhysRevB.47.558} {\bibfield  {journal}
  {\bibinfo  {journal} {Phys. Rev. B}\ }\textbf {\bibinfo {volume} {47}},\
  \bibinfo {pages} {558} (\bibinfo {year} {1993})}\BibitemShut {NoStop}%
\bibitem [{\citenamefont {Kresse}\ and\ \citenamefont
  {Furthm\"uller}(1996)}]{Kresse1996a}%
  \BibitemOpen
  \bibfield  {author} {\bibinfo {author} {\bibfnamefont {G.}~\bibnamefont
  {Kresse}}\ and\ \bibinfo {author} {\bibfnamefont {J.}~\bibnamefont
  {Furthm\"uller}},\ }\href {\doibase 10.1103/PhysRevB.54.11169} {\bibfield
  {journal} {\bibinfo  {journal} {Phys. Rev. B}\ }\textbf {\bibinfo {volume}
  {54}},\ \bibinfo {pages} {11169} (\bibinfo {year} {1996})}\BibitemShut
  {NoStop}%
\bibitem [{\citenamefont {Kresse}\ and\ \citenamefont
  {Furthmüller}(1996)}]{Kresse1996b}%
  \BibitemOpen
  \bibfield  {author} {\bibinfo {author} {\bibfnamefont {G.}~\bibnamefont
  {Kresse}}\ and\ \bibinfo {author} {\bibfnamefont {J.}~\bibnamefont
  {Furthmüller}},\ }\href {\doibase
  https://doi.org/10.1016/0927-0256(96)00008-0} {\bibfield  {journal} {\bibinfo
   {journal} {Computational Materials Science}\ }\textbf {\bibinfo {volume}
  {6}},\ \bibinfo {pages} {15 } (\bibinfo {year} {1996})}\BibitemShut {NoStop}%
\bibitem [{\citenamefont {Pustogow}\ \emph {et~al.}(2018)\citenamefont
  {Pustogow}, \citenamefont {Bories}, \citenamefont {L{\"{o}}hle},
  \citenamefont {R{\"{o}}sslhuber}, \citenamefont {Zhukova}, \citenamefont
  {Gorshunov}, \citenamefont {Tomi{\'{c}}}, \citenamefont {Schlueter},
  \citenamefont {H{\"{u}}bner}, \citenamefont {Hiramatsu}, \citenamefont
  {Yoshida}, \citenamefont {Saito}, \citenamefont {Kato}, \citenamefont {Lee},
  \citenamefont {Dobrosavljevi{\'{c}}}, \citenamefont {Fratini},\ and\
  \citenamefont {Dressel}}]{Pustogow2018}%
  \BibitemOpen
  \bibfield  {author} {\bibinfo {author} {\bibfnamefont {A.}~\bibnamefont
  {Pustogow}}, \bibinfo {author} {\bibfnamefont {M.}~\bibnamefont {Bories}},
  \bibinfo {author} {\bibfnamefont {A.}~\bibnamefont {L{\"{o}}hle}}, \bibinfo
  {author} {\bibfnamefont {R.}~\bibnamefont {R{\"{o}}sslhuber}}, \bibinfo
  {author} {\bibfnamefont {E.}~\bibnamefont {Zhukova}}, \bibinfo {author}
  {\bibfnamefont {B.}~\bibnamefont {Gorshunov}}, \bibinfo {author}
  {\bibfnamefont {S.}~\bibnamefont {Tomi{\'{c}}}}, \bibinfo {author}
  {\bibfnamefont {J.~A.}\ \bibnamefont {Schlueter}}, \bibinfo {author}
  {\bibfnamefont {R.}~\bibnamefont {H{\"{u}}bner}}, \bibinfo {author}
  {\bibfnamefont {T.}~\bibnamefont {Hiramatsu}}, \bibinfo {author}
  {\bibfnamefont {Y.}~\bibnamefont {Yoshida}}, \bibinfo {author} {\bibfnamefont
  {G.}~\bibnamefont {Saito}}, \bibinfo {author} {\bibfnamefont
  {R.}~\bibnamefont {Kato}}, \bibinfo {author} {\bibfnamefont {T.-H.}\
  \bibnamefont {Lee}}, \bibinfo {author} {\bibfnamefont {V.}~\bibnamefont
  {Dobrosavljevi{\'{c}}}}, \bibinfo {author} {\bibfnamefont {S.}~\bibnamefont
  {Fratini}}, \ and\ \bibinfo {author} {\bibfnamefont {M.}~\bibnamefont
  {Dressel}},\ }\href {\doibase 10.1038/s41563-018-0140-3} {\bibfield
  {journal} {\bibinfo  {journal} {Nat. Mater.}\ }\textbf {\bibinfo {volume}
  {17}},\ \bibinfo {pages} {773} (\bibinfo {year} {2018})}\BibitemShut
  {NoStop}%
\bibitem [{\citenamefont {Dressel}\ and\ \citenamefont
  {Pustogow}(2018)}]{Dressel2018}%
  \BibitemOpen
  \bibfield  {author} {\bibinfo {author} {\bibfnamefont {M.}~\bibnamefont
  {Dressel}}\ and\ \bibinfo {author} {\bibfnamefont {A.}~\bibnamefont
  {Pustogow}},\ }\href {http://stacks.iop.org/0953-8984/30/i=20/a=203001}
  {\bibfield  {journal} {\bibinfo  {journal} {J. Phys. Condens. Matter}\
  }\textbf {\bibinfo {volume} {30}},\ \bibinfo {pages} {203001} (\bibinfo
  {year} {2018})}\BibitemShut {NoStop}%
\bibitem [{\citenamefont {Ferber}\ \emph {et~al.}(2014)\citenamefont {Ferber},
  \citenamefont {Foyevtsova}, \citenamefont {Jeschke},\ and\ \citenamefont
  {Valent{\'{i}}}}]{Ferber2014}%
  \BibitemOpen
  \bibfield  {author} {\bibinfo {author} {\bibfnamefont {J.}~\bibnamefont
  {Ferber}}, \bibinfo {author} {\bibfnamefont {K.}~\bibnamefont {Foyevtsova}},
  \bibinfo {author} {\bibfnamefont {H.~O.}\ \bibnamefont {Jeschke}}, \ and\
  \bibinfo {author} {\bibfnamefont {R.}~\bibnamefont {Valent{\'{i}}}},\ }\href
  {https://link.aps.org/doi/10.1103/PhysRevB.89.205106} {\bibfield  {journal}
  {\bibinfo  {journal} {Phys. Rev. B}\ }\textbf {\bibinfo {volume} {89}},\
  \bibinfo {pages} {205106} (\bibinfo {year} {2014})}\BibitemShut {NoStop}%
\bibitem [{\citenamefont {Dressel}\ \emph {et~al.}(2016)\citenamefont
  {Dressel}, \citenamefont {Lazi{\'{c}}}, \citenamefont {Pustogow},
  \citenamefont {Zhukova}, \citenamefont {Gorshunov}, \citenamefont
  {Schlueter}, \citenamefont {Milat}, \citenamefont {Gumhalter},\ and\
  \citenamefont {Tomi{\'{c}}}}]{Dressel2016}%
  \BibitemOpen
  \bibfield  {author} {\bibinfo {author} {\bibfnamefont {M.}~\bibnamefont
  {Dressel}}, \bibinfo {author} {\bibfnamefont {P.}~\bibnamefont
  {Lazi{\'{c}}}}, \bibinfo {author} {\bibfnamefont {A.}~\bibnamefont
  {Pustogow}}, \bibinfo {author} {\bibfnamefont {E.}~\bibnamefont {Zhukova}},
  \bibinfo {author} {\bibfnamefont {B.}~\bibnamefont {Gorshunov}}, \bibinfo
  {author} {\bibfnamefont {J.~A.}\ \bibnamefont {Schlueter}}, \bibinfo {author}
  {\bibfnamefont {O.}~\bibnamefont {Milat}}, \bibinfo {author} {\bibfnamefont
  {B.}~\bibnamefont {Gumhalter}}, \ and\ \bibinfo {author} {\bibfnamefont
  {S.}~\bibnamefont {Tomi{\'{c}}}},\ }\href
  {https://link.aps.org/doi/10.1103/PhysRevB.93.081201} {\bibfield  {journal}
  {\bibinfo  {journal} {Phys. Rev. B}\ }\textbf {\bibinfo {volume} {93}},\
  \bibinfo {pages} {81201} (\bibinfo {year} {2016})}\BibitemShut {NoStop}%
\bibitem [{\citenamefont {Matsuura}\ \emph {et~al.}(2019)\citenamefont
  {Matsuura}, \citenamefont {Sasaki}, \citenamefont {Iguchi}, \citenamefont
  {Gati}, \citenamefont {M\"uller}, \citenamefont {Stockert}, \citenamefont
  {Piovano}, \citenamefont {B\"ohm}, \citenamefont {Park}, \citenamefont
  {Biswas}, \citenamefont {Winter}, \citenamefont {Valent\'{\i}}, \citenamefont
  {Nakao},\ and\ \citenamefont {Lang}}]{Matsuura2019}%
  \BibitemOpen
  \bibfield  {author} {\bibinfo {author} {\bibfnamefont {M.}~\bibnamefont
  {Matsuura}}, \bibinfo {author} {\bibfnamefont {T.}~\bibnamefont {Sasaki}},
  \bibinfo {author} {\bibfnamefont {S.}~\bibnamefont {Iguchi}}, \bibinfo
  {author} {\bibfnamefont {E.}~\bibnamefont {Gati}}, \bibinfo {author}
  {\bibfnamefont {J.}~\bibnamefont {M\"uller}}, \bibinfo {author}
  {\bibfnamefont {O.}~\bibnamefont {Stockert}}, \bibinfo {author}
  {\bibfnamefont {A.}~\bibnamefont {Piovano}}, \bibinfo {author} {\bibfnamefont
  {M.}~\bibnamefont {B\"ohm}}, \bibinfo {author} {\bibfnamefont {J.~T.}\
  \bibnamefont {Park}}, \bibinfo {author} {\bibfnamefont {S.}~\bibnamefont
  {Biswas}}, \bibinfo {author} {\bibfnamefont {S.~M.}\ \bibnamefont {Winter}},
  \bibinfo {author} {\bibfnamefont {R.}~\bibnamefont {Valent\'{\i}}}, \bibinfo
  {author} {\bibfnamefont {A.}~\bibnamefont {Nakao}}, \ and\ \bibinfo {author}
  {\bibfnamefont {M.}~\bibnamefont {Lang}},\ }\href {\doibase
  10.1103/PhysRevLett.123.027601} {\bibfield  {journal} {\bibinfo  {journal}
  {Phys. Rev. Lett.}\ }\textbf {\bibinfo {volume} {123}},\ \bibinfo {pages}
  {027601} (\bibinfo {year} {2019})}\BibitemShut {NoStop}%
\bibitem [{\citenamefont {Pilon}\ \emph {et~al.}(2013)\citenamefont {Pilon},
  \citenamefont {Lui}, \citenamefont {Han}, \citenamefont {Shrekenhamer},
  \citenamefont {Frenzel}, \citenamefont {Padilla}, \citenamefont {Lee},\ and\
  \citenamefont {Gedik}}]{Pilon2013}%
  \BibitemOpen
  \bibfield  {author} {\bibinfo {author} {\bibfnamefont {D.~V.}\ \bibnamefont
  {Pilon}}, \bibinfo {author} {\bibfnamefont {C.~H.}\ \bibnamefont {Lui}},
  \bibinfo {author} {\bibfnamefont {T.~H.}\ \bibnamefont {Han}}, \bibinfo
  {author} {\bibfnamefont {D.}~\bibnamefont {Shrekenhamer}}, \bibinfo {author}
  {\bibfnamefont {A.~J.}\ \bibnamefont {Frenzel}}, \bibinfo {author}
  {\bibfnamefont {W.~J.}\ \bibnamefont {Padilla}}, \bibinfo {author}
  {\bibfnamefont {Y.~S.}\ \bibnamefont {Lee}}, \ and\ \bibinfo {author}
  {\bibfnamefont {N.}~\bibnamefont {Gedik}},\ }\href
  {https://link.aps.org/doi/10.1103/PhysRevLett.111.127401} {\bibfield
  {journal} {\bibinfo  {journal} {Phys. Rev. Lett.}\ }\textbf {\bibinfo
  {volume} {111}},\ \bibinfo {pages} {127401} (\bibinfo {year}
  {2013})}\BibitemShut {NoStop}%
\bibitem [{Note1()}]{Note1}%
  \BibitemOpen
  \bibinfo {note} {See Supplemental Material which contains all 54 phonon modes
  for Herbertsmithite}\BibitemShut {NoStop}%
\bibitem [{\citenamefont {Kozlenko}\ \emph {et~al.}(2012)\citenamefont
  {Kozlenko}, \citenamefont {Kusmartseva}, \citenamefont {Lukin}, \citenamefont
  {Keen}, \citenamefont {Marshall}, \citenamefont {de~Vries},\ and\
  \citenamefont {Kamenev}}]{Kozlenko2012}%
  \BibitemOpen
  \bibfield  {author} {\bibinfo {author} {\bibfnamefont {D.~P.}\ \bibnamefont
  {Kozlenko}}, \bibinfo {author} {\bibfnamefont {A.~F.}\ \bibnamefont
  {Kusmartseva}}, \bibinfo {author} {\bibfnamefont {E.~V.}\ \bibnamefont
  {Lukin}}, \bibinfo {author} {\bibfnamefont {D.~A.}\ \bibnamefont {Keen}},
  \bibinfo {author} {\bibfnamefont {W.~G.}\ \bibnamefont {Marshall}}, \bibinfo
  {author} {\bibfnamefont {M.~A.}\ \bibnamefont {de~Vries}}, \ and\ \bibinfo
  {author} {\bibfnamefont {K.~V.}\ \bibnamefont {Kamenev}},\ }\href {\doibase
  10.1103/PhysRevLett.108.187207} {\bibfield  {journal} {\bibinfo  {journal}
  {Phys. Rev. Lett.}\ }\textbf {\bibinfo {volume} {108}},\ \bibinfo {pages}
  {187207} (\bibinfo {year} {2012})}\BibitemShut {NoStop}%
\bibitem [{\citenamefont {Biesner}\ and\ \citenamefont
  {Uykur}(2020)}]{Biesner2020}%
  \BibitemOpen
  \bibfield  {author} {\bibinfo {author} {\bibfnamefont {T.}~\bibnamefont
  {Biesner}}\ and\ \bibinfo {author} {\bibfnamefont {E.}~\bibnamefont
  {Uykur}},\ }\href {\doibase 10.3390/cryst10010004} {\enquote {\bibinfo
  {title} {{Pressure-Tuned Interactions in Frustrated Magnets: Pathway to
  Quantum Spin Liquids?}}}\ } (\bibinfo {year} {2020})\BibitemShut {NoStop}%
\bibitem [{\citenamefont {Wosnitza}\ \emph {et~al.}(2016)\citenamefont
  {Wosnitza}, \citenamefont {Zvyagin},\ and\ \citenamefont
  {Zherlitsyn}}]{Wosnitza2016}%
  \BibitemOpen
  \bibfield  {author} {\bibinfo {author} {\bibfnamefont {J.}~\bibnamefont
  {Wosnitza}}, \bibinfo {author} {\bibfnamefont {S.~A.}\ \bibnamefont
  {Zvyagin}}, \ and\ \bibinfo {author} {\bibfnamefont {S.}~\bibnamefont
  {Zherlitsyn}},\ }\href {\doibase 10.1088/0034-4885/79/7/074504} {\bibfield
  {journal} {\bibinfo  {journal} {Reports on Progress in Physics}\ }\textbf
  {\bibinfo {volume} {79}},\ \bibinfo {pages} {074504} (\bibinfo {year}
  {2016})}\BibitemShut {NoStop}%
\end{thebibliography}%

\clearpage
\newpage
\end{document}